\documentclass[pra,twocolumn]{revtex4} 
\usepackage{graphicx}
\usepackage{bm}
\usepackage{amsmath, amssymb}

\usepackage{color}
\begin{document}

\title{Stability of the Superfluid State in Three-Component 
Fermionic Optical Lattice Systems}
\author{Yuki Okanami}
\affiliation{Department of Physics, Tokyo Institute of Technology, Meguro-ku, Tokyo 152-8551, Japan}

\author{Nayuta Takemori}
\affiliation{Department of Physics, Tokyo Institute of Technology, Meguro-ku, Tokyo 152-8551, Japan}

\author{Akihisa Koga}
\affiliation{Department of Physics, Tokyo Institute of Technology, Meguro-ku, Tokyo 152-8551, Japan}

\date{\today}%

\begin{abstract}
Three-component fermionic
optical lattice systems are investigated 
in dynamical mean-field theory for the Hubbard model.
Solving the effective impurity model by means of 
continuous-time quantum Monte Carlo simulations in the Nambu formalism,
we find that the $s$-wave superfluid state proposed recently is 
indeed stabilized in the repulsively interacting case and 
appears along the first-order phase boundary 
between the metallic and paired Mott states in the paramagnetic system.
The BCS-BEC crossover in the three-component fermionic system
is also addressed.
\end{abstract}
\maketitle

\section{Introduction}
Ultracold atomic systems provide a variety of interesting topics 
\cite{ultracold}.
One of the most active topics is the superfluid (SF) state 
in ultracold fermions,
which has widely been investigated since the observation of the crossover 
between the Bardeen-Cooper-Schrieffer (BCS) and 
the Bose-Einstein condensation (BEC) states 
\cite{crossover1,crossover2,crossover3,crossover4}.
The high controllability of interaction strength, 
particle number, and other parameters enables us to study 
pseudogap behavior \cite{pseudo1,pseudo2} and
the SF state in other two-component fermionic systems 
such as the population-imbalanced and 
mass-imbalanced systems \cite{pop,mass,mass2}. 
Recently, degenerate multi-component fermionic systems 
have experimentally been realized \cite{Li,Li2,Yb,Yb2,Sr}, 
which stimulates further theoretical investigation 
on the SF state in multi-component fermionic systems. 

One of the simplest systems is the three-component fermionic optical lattice.
Ground-state properties of the system have been studied
in the Hubbard model 
where Mott transitions have been discussed
in the paramagnetic case \cite{Gorelik,Inaba1,Miyatake}.
Recently, it has been suggested that the $s$-wave SF ground state is realized 
when two of three on-site repulsive interactions are strong \cite{Inaba2}.
This is contrast to the fact that the SF state 
in the two-component Hubbard model is stabilized by attractive interactions.
Therefore, it is instructive to clarify whether or not 
the SF state realized in the three-component systems with repulsive interaction 
is adiabatically connected to that induced by the attractive interaction.
In addition, it may be important 
to examine normal-state properties in the three-component system
since the SF state is suggested to appear 
between the metallic and Mott states \cite{Inaba2}.
Hence, it is desired to study systematically particle correlations 
in the three-component interacting fermion systems. 

For this purpose, we consider the three-component Hubbard model,
combining dynamical mean-field theory 
(DMFT) \cite{Metzner,Muller,Georges,Pruschke} 
with a continuous-time quantum Monte Carlo (CTQMC) method \cite{solver_review,Werner}.
By calculating the pair potential, double occupancy, and 
renormalization factor, we determine finite-temperature phase diagrams 
of the system.
The BCS-BEC crossover in the SF state is also discussed.

The paper is organized as follows.
In Sec. II, we introduce the three-component Hubbard model and 
briefly summarize our theoretical approach.
In Sec. III, 
we study grand-state properties in the paramagnetic case 
to discuss the competition between the metallic and Mott states.
We then clarify the stability of the SF state 
at low temperatures in Sec. IV.
A brief summary is given in the last section.

\section{Model and Method}\label{2}
We consider three-component fermions in an optical lattice,
which should be described by the following Hubbard Hamiltonian,
\begin{equation}
\hat{\cal{H}}=-t\sum_{\langle i,j \rangle ,\alpha}
c^{\dagger}_{i\alpha}c_{j\alpha}
+\frac{1}{2}\sum_{\alpha \neq \beta,i} U_{\alpha \beta}n_{i\alpha}n_{i\beta},
\label{eq1}
\end{equation}
where $\langle i,j\rangle$ denotes 
the summation over the nearest neighbor sites,
$c_{i\alpha}$ ($c^{\dagger}_{i\alpha}$) is the annihilation (creation) operator
of a fermion with color $\alpha (=1,2,3)$ on the $i$th site, 
and $n_{i\alpha}= c^{\dagger}_{i\alpha}c_{i\alpha}$.
Here, $t$ is the transfer integral, 
and $U_{\alpha\beta}(=U_{\beta\alpha})$ is the on-site interaction 
between two fermions with colors $\alpha$ and $\beta$.
For simplicity, we set $U_{12}\equiv U$ and $U_{23}=U_{31}\equiv U'$ and
neglect the translational symmetry breaking phases such
as the density wave and magnetically ordered states~\cite{Inaba2}.
In the paper, we focus on the half-filled system with equally populations,
setting the chemical potential as 
$\mu_{\alpha} = \sum_{\beta\neq\alpha}U_{\alpha \beta}/2$.
We then consider the normal-state properties and 
the stability of the $s$-wave SF state.
In the latter case, we assume that Cooper pairs are
formed by the fermions with colors 1 and 2, and 
the pair potential $\Delta=\langle c_{i1}c_{i2}\rangle$ is regarded 
as an order parameter of the SF state.

First, we consider the symmetry of the Hamiltonian 
${\cal H}(t, U_{12}, U_{23}, U_{31})$ at half filling. 
When the lattice is bipartite, 
we can use the particle-hole transformations \cite{Shiba} as 
$c_{j1}  \rightarrow  (-1)^j c_{j1}^\dag ,
c_{j2}  \rightarrow  (-1)^j c_{j2}^\dag$,  
and $c_{j3} \rightarrow c_{j3}$ 
and obtain the Hamiltonian ${\cal H}(t, U_{12}, -U_{23}, -U_{31})$.
Therefore, our discussions can be restricted to the positive 
$U'(=U_{23}=U_{31})$ case
without loss of generality.
We also note that the pair potential defined above 
is invariant under the transformation.

An important point is that the Hubbard model eq. (\ref{eq1})
is reduced to interesting models
in some limits, where ground state properties have been discussed.
When $U=U'$, the system is reduced to the SU(3) Hubbard model,
where the stability of the metallic and SF states has been 
discussed\cite{Gorelik,Inaba1,Miyatake,Honerkamp,Rapp,Inaba3}. 
In the case $U'=0$, the system is divided into two systems: 
the two-component interacting fermions (colors 1 and 2)
and free fermions (color 3).
In the interacting fermions, 
a Mott transition occurs around $U/D\sim 3$~\cite{IPT} while 
a pairing transition occurs around $U/D\sim -3$~\cite{Capone}
in the paramagnetic state.
If one allows the ordered state with translational symmetry,
the SF ground state is always stabilized 
in the attractive case~\cite{Garg,Bauer,Toschi}.
When the interaction is anisotropic $(U\neq U')$,
the existence of the $s$-wave SF state in repulsively interacting case
has been clarified \cite{Inaba2}. 
This SF state is realized in a certain region 
between the metallic and Mott states, and thereby
it is necessary to treat the competing phases carefully. 
However, the treatment might be simple to discuss
the quantum phase transitions correctly, {\it e.g.}  
the iterative perturbation theory based on 
the second-order perturbation
sometimes underestimates particle correlations~\cite{IPT},  
and the single-site self-energy functional theory 
should be difficult to study 
the first-order transitions quantitatively~\cite{SFA}.
Furthermore, it is still unclear how this SF state 
is adiabatically connected  to 
the conventional SF state described by the BCS theory.
Therefore, it is desired to discuss systematically 
the stability of the SF state
in the three-component fermionic Hubbard model.

Before discussions, 
we consider the stability of the SF state by means of 
the simple mean-field theory.
When the static mean-field $\Delta$ is introduced,
the effective Hamiltonian is given by
$\hat{\cal H}_{BCS} = -t 
\sum_{\langle i,j \rangle ,\alpha}c^{\dagger}_{i\alpha}c_{j\alpha}
+U\sum_{i} \left(\Delta^*  c_{i 1} c_{i 2}  +h.c.\right). $
We note that the interaction $U'$ is irrelevant 
and fermions with color 3 play no role in stabilizing the SF state.
In the case, the $s$-wave SF state is 
stabilized only when the interaction is attractive $(U < 0)$.
On the other hand, in the repulsive case $(U>0)$, 
fermions with color 1 and 2 have no chance to form Cooper pairs, 
which is contrast to the previous work \cite{Inaba2}.
Furthermore, this theory is not appropriate to access 
the strong coupling region, where the Mott state should be stabilized. 
Therefore, it is necessary to incorporate 
particle correlations correctly.

To this end, we make use of DMFT \cite{Metzner,Muller,Georges,Pruschke}.
In DMFT, the original lattice model is mapped to an effective impurity model, 
where local and dynamical correlations can be taken into account.
The lattice Green's function is obtained via a self-consistency condition 
imposed on the impurity problem. 
This allows us to discuss the stability of the $s$-wave SF state 
more quantitatively beyond the static BCS mean-field 
theory \cite{GeorgesZ}.
In fact, the DMFT method has successfully been applied to 
various strongly correlated fermion systems with 
the SF or superconducting 
states~\cite{Garg,Dao,Haule,Takemori,Bodensiek,KogaSF,KogaQMC1,Hoshino}.

When the SF state is treated in the framework of DMFT,
the impurity Green's function for the effective model 
$\hat{G}_{\rm imp}(\tau)$ should be described by a $3\times 3$ matrix as,
\begin{align}
\hat{G}_{\rm{imp}}(\tau) =
\begin{pmatrix}
G_{1}(\tau) & F(\tau) & 0 \\
F^*(\tau) & -G_{2}(-\tau) & 0 \\
0 & 0 & G_{3}(\tau)
\end{pmatrix},
\end{align}
where $G_\alpha(\tau)=-\langle T_\tau f_\alpha(\tau) f_\alpha^\dag(0) \rangle$
denotes the normal Green's function for color $\alpha$, and 
$F(\tau)=-\langle T_\tau f_1(\tau) f_2(0)\rangle$ and 
$F^*(\tau)=-\langle T_\tau f_2^\dag(\tau) f_1^\dag(0)\rangle$
denote the anomalous Green's functions.
In the calculations, we use a semi-circular density of states,
$\rho(x) = 2 /(\pi D) \sqrt{1-(x/D)^2}$,
where $D$ is the half bandwidth. 
The self-consistency equation is given by
\begin{align}
\hat{G}_{0,\rm imp}^{-1}&(i\omega_n) = i\omega_n \hat{1} +
\hat{\mu} - \left( \frac{D}{2} \right) ^{2} \hat{\Lambda} \hat{G}_{\rm imp}(i\omega_{n}) \hat{\Lambda}, 
\end{align}
where $\hat{1}$ is the identity matrix, 
$\hat\mu={\rm diag}(\mu_1, -\mu_2, \mu_3)$, 
$\hat\Lambda={\rm diag}(1,-1,1)$, 
$\omega_n =(2n+1)\pi T$ is the Matsubara frequency, and
$T$ is the temperature. 
Here, $\hat{G}_{0,\rm imp}$ and $\hat{G}_{\rm imp}$ are noninteracting 
and full Green's functions for the effective impurity model.

There are various methods to solve the effective impurity problem.
To study the stability of the SF state 
in a three-component fermionic optical lattice system, 
an unbiased and accurate numerical solver is necessary
such as an exact diagonalization~\cite{Caffarel}
and the numerical renormalization group~\cite{NRG,NRG_RMP}.
A particularly powerful method for exploring finite-temperature properties 
is the hybridization-expansion CTQMC method~\cite{Werner,solver_review}.
This enables us to study the Hubbard model 
in both weak- and strong-coupling regimes.

In this paper, we use the half bandwidth $D$ as a unit of energy. 
We calculate the double occupancy 
for colors $\alpha$ and $\beta$
$D_{\alpha \beta}$, pair potential $\Delta$,
internal energy $E$, and specific heat $C$, which are given by 
\begin{eqnarray}
D_{\alpha \beta}&=&\langle n_{i\alpha} n_{i\beta} \rangle,\\
\Delta&=&\langle c_{i1}c_{i2}\rangle=\lim_{\tau\rightarrow 0_+} F(\tau),\\
E&=&\left(\frac{D}{2}\right)^2\int_0^\beta d\tau 
\rm{Tr}[ \hat{G}_{imp}(\tau) \hat{\Lambda} \hat{G}_{imp} (-\tau) \hat{\Lambda} ]   \nonumber \\
&+&\frac{1}{2}\sum_{\alpha\beta} U_{\alpha\beta}D_{\alpha\beta},\\
C&=&\frac{dE}{dT}.
\end{eqnarray}
We also calculate the quantity $Z_\alpha=[1-{\rm Im} 
\Sigma_\alpha(i\omega_0)/\omega_0]^{-1}$ as the quasi-particle weight 
at finite temperatures,
where $\Sigma_{\alpha}$ is the normal self-energy for color $\alpha$. 
In addition to these static quantities, we compute the density
of states, applying the maximum entropy method (MEM) \cite{MEM1,MEM2,MEM3}
to the normal Green's function. We then discuss how the gap
structure appears around the Fermi level at low temperatures.

\section{Low-temperature properties in the paramagnetic state}\label{3}
\begin{figure}[htb]
\begin{center}
\includegraphics[width=8cm]{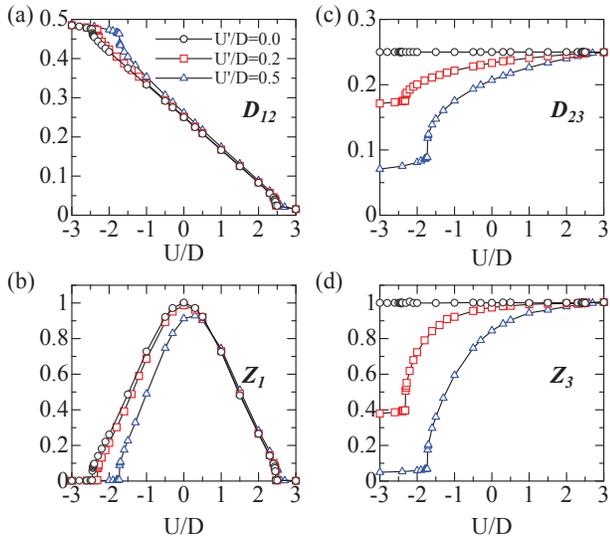}
\caption{(Color online).
Double occupancies $D_{\alpha\beta}$ and 
the renomalization factors $Z_\alpha$
as functions of $U/D$ at $T/D=0.015$.
Open circles, squares and triangles represent the results
for the half-filled system with $U'/D=0.0$, $0.2$ and $0.5$.
}
\label{fig:paradz}
\end{center}
\end{figure}

In the section,
we consider low-temperature properties of the half-filled system 
in the paramagnetic state.
Then we study quantitatively how the Mott and pairing transitions 
in the two-component Hubbard model are connected 
to the phase transitions in the three-component systems~\cite{Inaba1,Miyatake}.
By performing DMFT with CTQMC method, we calculate
the double occupancies $D_{\alpha\beta}$ 
and renormalization factors $Z_\alpha$ 
at the temperature $T/D=0.015$, as shown in Fig. \ref{fig:paradz}.
First, we consider the system with $U'=0$, which is divided into two systems:
two-component interacting fermions (colors 1 and 2) 
and free fermions (color 3).
Now, we focus on the two-component fermions.
As decreasing interaction $U$ from zero,
we find that 
the double occupancy $D_{12}$ is smoothly increased and 
the renormalization factor $Z_1$ is decreased 
as shown in Figs. \ref{fig:paradz} (a) and (b). 
In the weak-coupling region $(U_p<U)$, the metallic state is realized with a
finite renormalization factor,
where $U_p/D\sim -2.6$.
At $U=U_p$, the jump singularity appears in both curves and
the first-order phase transition occurs.
When $U<U_p$, each site is doubly occupied or empty ($D_{12}\sim 0.5$),
and the state is specified by the pairing state \cite{Capone}.
On the other hand, the increase in the interaction $U (>0)$ decreases
both $D_{12}$ and $Z_1$ 
and induces the phase transition at $U_m/D\sim 2.6$.
When $U_m<U$, each site is singly occupied ($D_{12}\sim 0.0$),
which implies that the Mott insulating state is realized.
We also find that these quantities are symmetric at $U=0.0$,
which is reflected by the fact that the repulsive Hubbard model is equivalent 
to the attractive one under the particle-hole transformation \cite{Shiba}.
In contrast to the two-component system, 
fermions with color 3 are always noninteracting, 
and thereby $D_{23}=0.25$ and $Z_3=1.0$.

\begin{figure}[htb]
\begin{center}
\includegraphics[width=7cm]{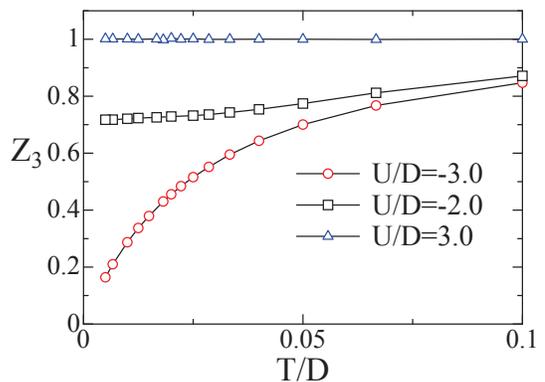}
\caption{(Color online) 
Open circles, squares and triangles represent the renormalization factor $Z_3$
in the system with a fixed $U'/D=0.2$ when $U/D=-3.0$, $-2.0$ and $3.0$.
}
\label{fig:para_Z_T}
\end{center}
\end{figure}

The introduction of the interaction $U'$ leads to a drastic change
in fermions with color 3, in particular, when $U<0$.
A clear jump singularity appears in the curves of
the double occupancy $D_{23}$ and the renormalization factor $Z_{3}$ 
at $U_p/D\sim -2.4$ while a tiny one appears around $U_m/D\sim 2.6$
in the case $U'/D=0.2$.
When $U_p<U<U_m$, metallic behavior still remains 
with finite renormalization factors $Z_\alpha$ ($\alpha=1,2,3$).
In the large $U(>U_m)$ region, we find 
that $D_{12}\sim 0$, $D_{23}=D_{31}\sim 0.25$, $Z_1=Z_2\sim 0.0$ and $Z_3\sim 1.0$.
This implies that the Mott insulating state is realized for colors 1 and 2, 
and free fermion behavior survives for color 3.
Therefore, this state can be regarded as the color selective Mott (CSM) state \cite{Inaba1}.
By contrast, when $U<U_p$, 
$D_{23}$ and $Z_3$ are rapidly decreased 
on the introduction of the interaction $U'$,
as shown in Figs. \ref{fig:paradz} (c) and (d).
This suggests that metallic behavior for fermions with color 3 
is no longer stable at zero temperature although 
the renormalization factor $Z_3$ is not so small at $T/D=0.015$.
To confirm this, we also calculate the temperature dependence of
the renormalization factor $Z_3$, as shown in Fig. \ref{fig:para_Z_T}.
It is found that in the system with $U/D=-3.0$ and $U'/D=0.2$,
the renormalization factor decreases with lowering temperatures
and does not converge down to $T/D=0.005$.
This tendency suggests that the paired Mott (PM) ground state \cite{Inaba1}
is realized with $Z_\alpha\sim0.0$, $D_{12}\sim 0.5$, and $D_{23}\sim 0.0$ and
bad metallic behavior appears at intermediate temperatures.
On the other hand, in the case $U/D=-2.0$ ($U/D=3.0$), 
the renormalization factor $Z_3$ becomes constant at 
low temperatures, which is consistent with the fact that 
the metallic (CSM) ground state is realized.

\begin{figure}[htb]
\begin{center}
\includegraphics[width=8cm]{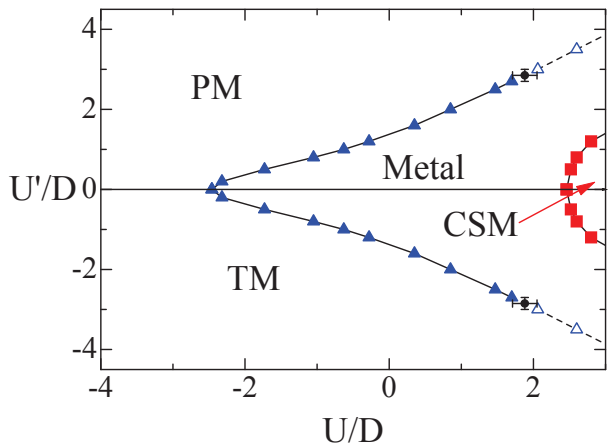}
\caption{(Color online) 
Phase diagram at $T/D = 0.015$ in the paramagnetic system.
Squares (triangles) represent the first-order phase transition points 
between the metallic and CSM (PM or TM) states, and Circles indicate
the critical end points.
Dash lines represent the crossover between the PM (TM) and metallic states.
}
\label{fig:PD1}
\end{center}
\end{figure}

By performing similar calculations, 
we obtain the phase diagram for the paramagnetic state at $T/D=0.015$, 
as shown in Fig. \ref{fig:PD1}. 
We find that four quantum states compete with each other,
such as CSM, metallic, PM and trion Mott (TM) states. 
Last two states are equivalent under the particle-hole transformations
and the TM state is realized only when $U'$ is negative.
In the region with $U/D\lesssim -2.6$ and small $|U'|$, 
fermions with colors 1 and 2 strongly
couple with each other due to the attractive interaction, and
fermions with color 3 weakly couple with the other fermions.
Therefore, bad metallic behavior appears at the temperature, 
as discussed above.
This state is adiabatically connected to the PM or TM state. 
Since the PM and TM states are stabilized by the interaction strength $|U'|$ 
at low temperatures,
the first-order phase transition point $U_p$ is rapidly increased, 
as shown in Fig. \ref{fig:PD1}.
In the large $U'$ region, the singularity in the physical quantities
at the phase boundary smears. 
We find the critical end point around $(U/D, U'/D)\sim (1.9, 2.9)$ .
Beyond the point, one finds no singularities and
the crossover between the PM and metallic states occurs.
By contrast, the other transition point $U_m$ is 
insensitive to the interaction $U'$.
This may originate from the nature of the CSM state.
In the state, a fermion with color 1 or 2 is singly occupied at each site, 
and fermions with color 3 are almost free.
Therefore, the introduction of the interaction $U'$
little affects low-temperature properties in the system.

In the section, we have discussed the competition between the CSM, metallic, 
PM, and TM states in the paramagnetic system.
In the following, we study how the $s$-wave SF state is 
realized between the metallic
and PM states and how the BCS-BEC crossover occurs in the system.

\section{Stability of  the superfluid state}\label{4}
\begin{figure}[htb]
\begin{center}
\includegraphics[width=8cm]{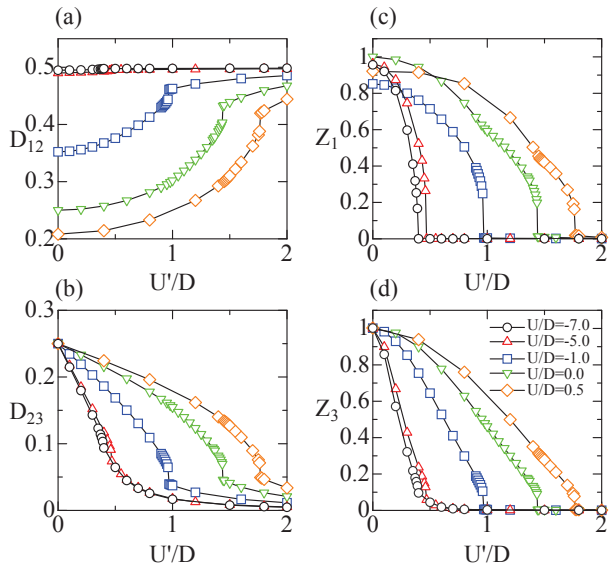}
\caption{(Color online) 
Double occupancies $D_{\alpha \beta}$ and 
renormalization factors $Z_{\alpha}$ 
at $T/D=0.015$ when $U/D=-7.0$, $-5.0$, $-1.0$, $0.0$, and $0.5$.
}
\label{fig:dzsuperUfix}
\end{center}
\end{figure}

In this section, we perform DMFT in the Nambu formalism
to discuss how the SF state is realized in the system at low temperatures.
When $U'=0.0$, 
the system is reduced to the two-component Hubbard model, where
the SF state is realized in the case ($-18 <U/D<-0.55$) 
at $T/D=0.015$. 
It is known that the BCS-BEC crossover, which is roughly characterized by 
the maximum of the pair potential, occurs around $U/D\sim -5.0$~\cite{KogaSF}.
\begin{figure}[htb]
\begin{center}
\includegraphics[width=7cm]{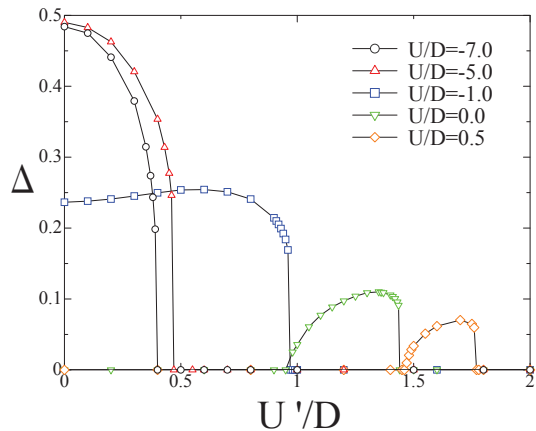}
\caption{(Color online) 
Pair potential $\Delta$ as functions of $U'/D$ in the system at $T/D=0.015$
when $U/D=-7.0$, $-5.0$, $-1.0$, $0.0$, and $0.5$. 
}
\label{fig:deltaU2}
\end{center}
\end{figure}
When the system is in the strong-coupling BEC region ($U/D=-7.0$, $U'=0.0$),  
the introduction of the interaction $U'$ monotonically decreases
the pair potential, and
finally it suddenly vanishes at $U'/D=0.4$,
as shown in Fig. \ref{fig:deltaU2}.
Jumps are also found in the curves of 
the double occupancy and renormalization factor 
although the singularities may be invisible in Fig. \ref{fig:dzsuperUfix}.
Therefore,
we can say that in the BEC region,
the first-order phase transition occurs to the PM state.
Similar behavior is also found in the BCS-BEC crossover region ($U/D=-5.0$, $U'=0.0$),
where the first-order phase transition occurs at $U'/D=0.47$.
On the other hand, in the weak-coupling BCS region ($U/D=-1.0$, $U'=0.0$), 
different behavior appears in the pair potential, 
as shown in Fig. \ref{fig:deltaU2}, although we can not find
an obvious change in the other quantities as shown in Fig. \ref{fig:dzsuperUfix}.
Namely, the pair potential has a maximum around $U'/D\sim 0.6$, 
as shown in Fig. \ref{fig:deltaU2}. 
This implies that the introduction of the interaction $U'$ stabilizes
the SF state and induces the BCS-BEC crossover.
Then, the first-order phase transition occurs 
from the BEC-type SF state to the PM state at $U'/D=0.97$.
This nonmonotonic behavior is more clearly found in the positive $U$ case.
When $U'=0$, the SF state is no longer realized.
As increasing $U'$, the metallic state becomes unstable
and the second-order phase transition occurs to the BCS-type SF state 
with the pair potential.
By examining critical behavior $\Delta\sim |U-U_c|^\beta$ with 
the exponent $\beta=1/2$, we determine the critical interactions 
$U'/D\sim 0.98$ ($U=0.0$) and $U'/D\sim 1.47$ ($U/D=0.5$) at $T/D=0.015$.
Further increase of $U'$ stabilizes the SF state, and
finally the first-order phase transition occurs to the PM state 
at $U'/D=1.44$ ($U=0.0$) and $U'/D=1.77$ ($U/D=0.5$).
An important point is that the SF state is realized even when 
all on-site interactions are repulsive.
These indicate that the interaction $U'$ plays an essential role 
in stabilizing the SF state in the repulsively interacting case
since the $s$-wave SF state does not appear 
in the two-component repulsive Hubbard model.

\begin{figure}[htb]
\begin{center}
\includegraphics[width=8cm]{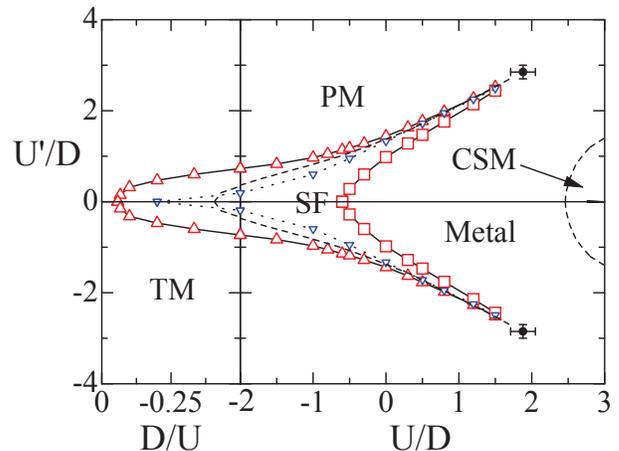}
\caption{(Color online) 
Phase diagram of the system at $T/D=0.015$. 
Triangles (squares) represent the phase transition points between 
the SF and PM (metallic) states.
The dashed lines represent the phase boundaries in the paramagnetic system. 
Dotted lines indicate the ridges of the pair potential in the SF state. 
(please delete inverted triangles)
}
\label{fig:PD2}
\end{center}
\end{figure}

By performing similar calculations for different values of $U$ and $U'$,
we obtain the phase diagram at the temperature $T/D=0.015$ 
as shown in Fig. \ref{fig:PD2}.
When $U'=0.0$, the system is reduced to the two-component Hubbard 
model for colors 1 and 2. 
In the BEC (strong-coupling) region, the introduction of the interaction $U'$
drives the system to the PM state, 
and its phase transition is of first order.
On the other hand, in the BCS (weak-coupling) region, 
the phase transition between the SF and metallic states is of second order.
A remarkable point is that the SF state emerges along the phase boundary 
between the metallic and PM states discussed in the previous section
and persists up to fairly large repulsive interactions.
Therefore, we can say that the SF state in the repulsive case
is adiabatically connected to the trivial SF state realized 
in the two-component attractive Hubbard model.
On the other hand, once the parameters $(U, U')$ are away from the line, 
the SF state becomes immediately unstable.
These mean that the SF state in the repulsively interacting case 
is induced by the competition between the on-site interactions $U$ and $U'$.
In the stronger coupling region, the SF state is no longer stable, and
the phase diagram is reduced to the paramagnetic one, where
the PM, metallic, and CSM states compete with each other 
(see Fig. \ref{fig:PD1}).
We wish to note that the critical end point for the SF state
corresponds to that in the paramagnetic phase diagram 
within our numerical accuracy.
Therefore, we can say that the first-order phase transition between
the PM and metallic states in the paramagnetic system plays
an essential role in stabilizing the SF state.
This is different from the origin of the weak-coupling SF state
discussed in terms of the random phase approximation~\cite{Inaba2}.
Although we have determined the phase diagram at the temperature $T/D=0.015$,
these results suggest that the SF state is stabilized in the stronger coupling
region $(U'\gtrsim U \gg D)$ at lower temperatures.

\begin{figure}[htb]
\begin{center}
\includegraphics[width=9cm]{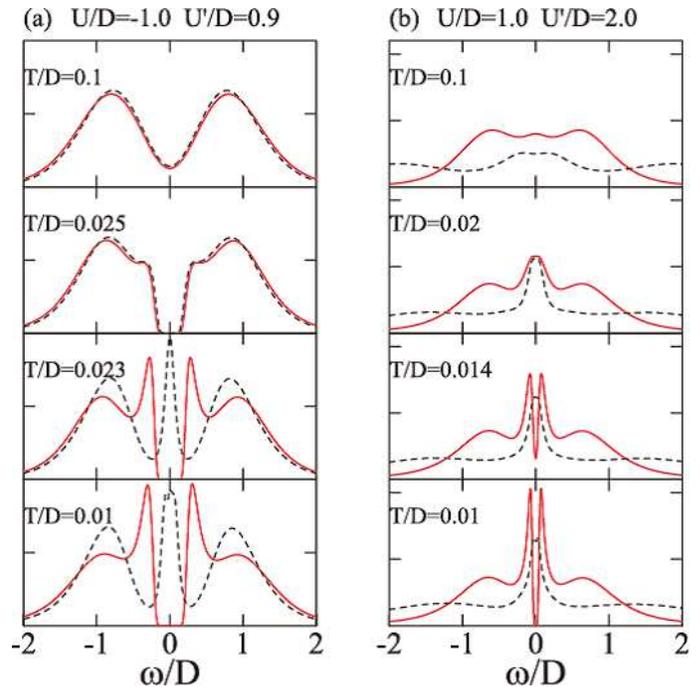}
\caption{(Color online) 
Solid (dashed) lines represent the density of states 
for fermions with colors 1 (color 3)
when $U/D=-1.0$ and $U'/D=0.9$ (a), and $U/D=1.0$ and $U'/D=2.0$ (b).}
\label{fig:BEC-dos}
\end{center}
\end{figure}

We also consider the BCS-BEC crossover in the three-component system.
This crossover should occur around the ridges of the pair potential 
in the parameter space $(U, U')$, except for the $U'=0.0$ axis.
Here, we examine low-temperature properties characteristic of 
the BCS and BEC regions.
The system with $U/D=-1.0$ and $U'/D=0.9$ 
belongs to the BEC-type SF state close to the first-order phase boundary
at $T/D=0.015$.
\begin{figure}[htb]
\begin{center}
\includegraphics[width=7cm]{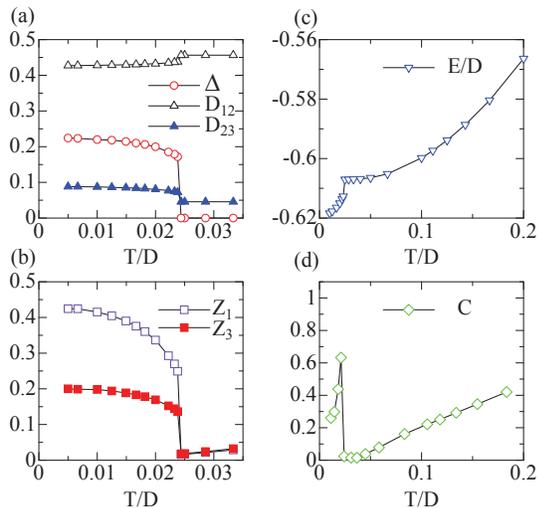}
\caption{(Color online)
The temperature dependence of the pair potential, double occupancies, 
renormalization factors, internal energy, and specific heat
when $U/D=-1.0$ and $U'/D =0.9$. 
}
\label{fig:BEC}
\end{center}
\end{figure}
The temperature dependence of physical quantities 
are shown in Fig. \ref{fig:BEC}. 
We find a jump singularity in each curve at the temperature $T_c/D =0.024$,
which means the existence of the first-order phase transition.
Above the temperature, double occupancies and renormalization factors are
calculated as $D_{12}\sim 0.5$, $D_{23}\sim 0.0$, and $Z_\alpha\sim 0.0$,
and the PM state is indeed stabilized.
The entropy for the PM state should be larger than that for the SF state 
at $T=T_c$, yielding
the discontinuity in the curve of the internal energy.
Therefore, we can find that the peak in the specific heat 
below the critical temperature is somewhat smaller than
that expected in the conventional $s$-wave SF state.
We also deduce the density of states by means of the MEM~\cite{MEM1,MEM2,MEM3},
as shown in Fig. \ref{fig:BEC-dos} (a).
In the BEC region, paired particles are formed even above 
the transition temperature, stabilizing the PM state.
Therefore, the large gap structure, 
which is proportional to the strength of the on-site interactions,
appears in the density of states for each color.
The sudden increase in the pair potential leads to
interesting behavior in dynamical properties.
The large gap and sharp peaks at its edges appear
in the density of states for colors 1 and 2.
On the other hand, the sharp peak structure is induced 
at the Fermi level in the density of states for color 3,
as shown in Fig. \ref{fig:BEC-dos} (a).
Therefore, below the transition temperature, 
paired particles are condensed, and 
the heavy metallic behavior appears for fermions with color 3.
Since the phase transition is accompanied with the insulator-metal transition
for fermions with color 3,
low-temperature properties are different from those
in the BEC region of the two-component system, where
the second-order phase transition occurs~\cite{KogaSF}.
Therefore, we can say that the BEC region in the three-component system 
should be characteristic of the first-order phase transition.

\begin{figure}[htb]
\begin{center}
\includegraphics[width=8cm]{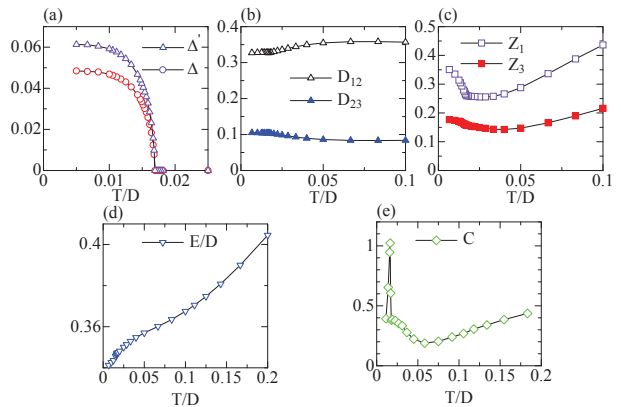}
\caption{(Color online) 
Temperature dependence of the pair potential, double occupancies, 
renormalization factors, 
internal energy, and specific heat when $U/D=1.0$ and $U'/D=2.0$.
}
\label{fig:delta2}
\end{center}
\end{figure}

In the BCS case ($U/D=1.0$ and $U'/D=2.0$),
different behavior appears, as shown in Fig. \ref{fig:delta2}.
We find that the decrease of the temperature induces the second-order 
phase transition.
The critical temperature is deduced as $T/D \sim0.017$, by examining 
critical behavior of the pair potential $\Delta \sim |T-T_c|^\beta$.
There is no discontinuity in the curve of the internal energy and
the large specific heat appears below the critical temperature,
in contrast to the BEC-type case.
We clearly find the local minimum in the specific heat around $T^*/D\sim 0.05$.
This behavior may be explained by the following.
At the higher temperature $T>T^*$,
the strength of the interaction affects low-energy properties and
gap behavior appears, where the renormalization factors
are decreased on the decrease of the temperature.
Around $T = T^*$, the competition between on-site interactions
begins to form the quasiparticles around the Fermi level,
giving rise to the minimum of the renormalization factors and specific heat,
as shown in Figs. \ref{fig:delta2} (c) and (e).
This is more clearly found in the density of states,
as shown in Fig. \ref{fig:BEC-dos} (b).
At high temperatures, gap behavior appears in the density of states 
for each color 
and the quasiparticle peaks develop on the decreasing temperatures.
Once the system enters the SF state, in the spectral functions 
with colors 1 and 2, a dip structure develops 
at the Fermi level while a sharp quasiparticle peak remains in the other.
These low-temperature properties are contrast to those in the BEC region.

When the parameters $(U, U')$ are varied in the SF state, 
the crossover occurs between 
the BCS-type and BEC-type SF states, and
low-temperature properties are gradually changed.
We find that the phase transition from the BEC-type (BCS-type) SF state 
to the paramagnetic state is of first (second) order.

\begin{figure}[htb]
\begin{center}
\includegraphics[width=7cm]{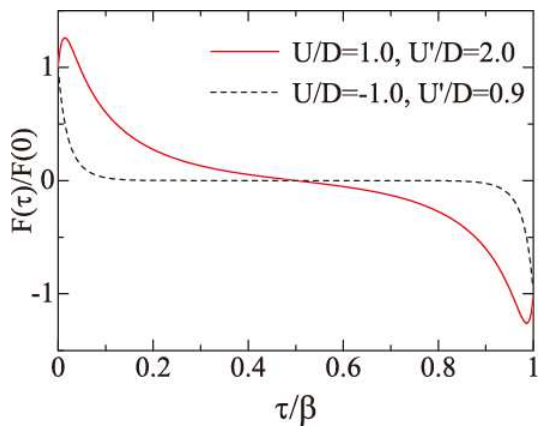}
\caption{(Color online)
Solid (dashed) line represents the normalized anomalous green's function at $T/D=0.01$
when $U/D=1.0, U'/D=2.0$ ($U/D=-1.0, U'/D=0.9$). 
}
\label{fig:F}
\end{center}
\end{figure}

Before closing this section, we briefly discuss 
the effect of the repulsive interaction on the SF state.
When $U/D=1.0$ and $U'/D=2.0$, the anomalous Green's function $F(\tau)$
at the low temperature $T/D=0.01$ is shown in Fig. \ref{fig:F}. 
We find that the anomalous Green's function takes its maximum value $\Delta'$ 
away from $\tau=0$.
This is contrast to the conventional anomalous Green's function 
in the SF state,
where the maximum is located at $\tau=0$ and its value can be regarded as 
the order parameter, as shown in Fig. \ref{fig:F}.
This may originate from the fact that 
the repulsive interaction between fermions with colors 1 and 2 
tends to break the Cooper pair at each site.
To clarify how such a nontrivial structure in $F(\tau)$ appears 
below the critical temperature,
we also show the temperature dependence of 
the maximum value $\Delta'$ in Fig. \ref{fig:delta2} (a).
As decreasing temperature, $\Delta$ and $\Delta'$ are simultaneously induced
at the critical temperature and are gradually increased at lower temperatures.
Since the ratio $\Delta'/\Delta \sim 1.3$ little depends on the temperature,
the SF state without the pair potential [$F(\tau)\neq 0, \Delta=0$] 
should not be realized at any temperatures in the three-component system.

\section{Conclusions}\label{4}
We have investigated the three-component Hubbard model 
in infinite dimensions, combining DMFT with the CTQMC method.
Computing the double occupancy and renormalization factor, 
we have determined the finite-temperature phase diagram 
in the paramagnetic state.
It has been clarified that the PM, CSM, TM, and metallic states 
compete with each other in the system.
We have also studied the stability of the SF state 
and have confirmed that the SF state is indeed stabilized 
in the repulsively interacting case. 
Systematic calculations have revealed that 
this state is adiabatically connected to 
the SF state realized in the two-component attractive Hubbard model.
Moreover, we have studied low-temperature properties 
in the BCS and BEC regions.
It is found that the BCS state is characterized by 
the second-order phase transition,
while the BEC state is by the first-order one.
This is contrast to the two-component system, 
where the second-order phase transitions occur in both limits.
It is also interesting how the SF state is realized in the multi-component 
fermionic systems for lithium and ytterbium atoms,
which is now under consideration.

\section*{Acknowledgments}
The authors thank K. Inaba and S. Suga for valuable discussions. 
This work was partly supported by the Grant-in-Aid for Scientific Research 
20740194 (A.K.) and 
the Global COE Program "Nanoscience and Quantum Physics" from 
the Ministry of Education, Culture, Sports, Science and Technology (MEXT) 
of Japan. 
N.T. also acknowledges the financial support from the Global Center of 
Excellence Program by MEXT, Japan 
through the "Nanoscience and Quantum Physics" 
Project of the Tokyo institute of Technology.
The simulations have been performed using some of 
the ALPS libraries~\cite{alps1.3}.


\begin{thebibliography}{99}
\bibitem{ultracold}
I. Bloch, J. Dalibard and S. Nascimb$\grave{e}$ne, 
Nat. Phys. {\bf 8}, 267 (2012).

\bibitem{crossover1}
C. A. Regal, M. Greiner, and D. S. Jin, 
Phys. Rev. Lett. {\bf 92}, 040403 (2004).

\bibitem{crossover2}
M. Bartenstein, A. Altmeyer, S. Riedl, S. Jochim, 
C. Chin, J. Hecker Denschlag, and R. Grimm, 
Phys. Rev. Lett. {\bf 92}, 120401 (2004). 

\bibitem{crossover3}
M. W. Zwierlein, C. A. Stan, C. H. Schunck, S. M. F. Raupach, 
A. J. Kerman, and W. Ketterle, Phys. Rev. Lett. {\bf 92}, 120403 (2004).

\bibitem{crossover4}
J. Kinast, S. L. Hemmer, M. E. Gehm, A. Turlapov, and J. E. Thomas, 
Phys. Rev. Lett. {\bf 92}, 150402 (2004). 

\bibitem{pseudo1}
J. T. Stewart, J. P. Gaebler, and D. S. Jin,
Nature (London) {\bf 454}, 744 (2008).

\bibitem{pseudo2}
J. P. Gaebler, J. T. Stewart, T. E. Drake, D. S. Jin, A. Perali, 
P. Pieri, and G. C. Strinati, 
Nat. Phys. {\bf 6}, 569 (2010).

\bibitem{pop}
M. W. Zwierlein, A. Schirotzek, C. H. Schunck, and W. Ketterle, 
Science {\bf 311}, 492 (2006).

\bibitem{mass}
M. Taglieber, A. -C. Voigt, T. Aoki, T. W. H$\ddot{a}$nsch, and K. Dieckmann, 
Phys. Rev. Lett. {\bf 100} 010401 (2008).

\bibitem{mass2}
H. Hara, Y. Takasu, Y. Yamaoka, J. M. Doyle, and Y. Takahashi, 
Phys. Rev. Lett. {\bf 106} 205304 (2011).

\bibitem{Li}
T. B. Ottenstein, T. Lompe, M. Kohnen, A. N. Wenz, and S. Jochim, 
Phys. Rev. Lett. {\bf 101} 203202 (2008).

\bibitem{Li2}
J. H. Huckans, J. R. Williams, E. L. Hazlett, R. W. Stites, and K. M. O'Hara, 
Phys. Rev. Lett. {\bf 102} 165302 (2009).

\bibitem{Yb}
T. Fukuhara, Y. Takasu, M. Kumakura, and Y. Takahashi, 
Phys. Rev. Lett. {\bf 98}, 030401 (2007).

\bibitem{Yb2}
S. Taie, Y. Takasu, S. Sugawa, R. Yamazaki, T. Tsujimoto, 
R. Murakami, and Y. Takahashi, 
Phys. Rev. Lett. {\bf 105}, 190401 (2010).

\bibitem{Sr}
B. J. DeSalvo, M. Yan, P. G. Mickelson, Y. N. Martinez de Escobar, and T. C. Killian, 
Phys. Rev. Lett. {\bf 105}, 030402 (2010).

\bibitem{Gorelik}
E. V. Gorelik and N. Bl\"umer, Phys. Rev. A {\bf 80}, 051602(R) (2009).

\bibitem{Inaba1}
K. Inaba, S. Miyatake, and S. Suga, 
Phys. Rev. A {\bf 82}, 051602(R) (2010).

\bibitem{Miyatake}
S. Miyatake, K. Inaba, and S. Suga, 
Phys. Rev. A {\bf 81}, 021603(R) (2010).

\bibitem{Inaba2}
K. Inaba and S. Suga, 
Phys. Rev. Lett. {\bf 108}, 255301 (2012);
K. Inaba and S. Suga, 
Mod. Phys. Lett B {\bf 27} 1330008 (2013).

\bibitem{Metzner}
W. Metzner and D. Vollhardt, 
Phys. Rev. Lett. {\bf 62}, 324 (1989).

\bibitem{Muller}
E. M\"uller-Hartmann, 
Z. Phys. B {\bf 74}, 507 (1989).

\bibitem{Georges}
A. Georges, G. Kotliar, W. Krauth, and M. J. Rozenberg,
Rev. Mod. Phys. {\bf 68}, 13 (1996).

\bibitem{Pruschke}
T. Pruschke, M. Jarrell, and J. K. Freericks,
Adv. Phys. {\bf 44}, 187 (1995).
\bibitem{solver_review}
E. Gull, A. J. Millis, A. N. Rubtsov, A. I. Lichtenstein, M. Troyer, and P. Werner,  
Rev. Mod. Phys. {\bf 83}, 349 (2011).

\bibitem{Werner}
P. Werner, A. Comanac, L. de’ Medici, M. Troyer, and A. J. Millis, 
Phys. Rev. Lett. {\bf 97}, 076405 (2006).

\bibitem{Shiba}
H. Shiba, 
Prog. Theor. Phys. {\bf 48}, 2171 (1972).

\bibitem{Honerkamp}
C. Honerkamp and W. Hofstetter, 
Phys. Rev. Lett. {\bf 92}, 170403 (2004); 
C. Honerkamp and W. Hofstetter, 
Phys. Rev. B {\bf 70}, 094521 (2004).

\bibitem{Rapp}
A. Rapp, G. Zar$\acute{a}$nd, C. Honerkamp, and W. Hofstetter, 
Phys. Rev. Lett. {\bf 98}, 160405 (2007); 
A. Rapp, W. Hofstetter, and G. Zar$\acute{a}$nd, 
Phys. Rev. B {\bf 77}, 144520 (2008). 

\bibitem{Inaba3}
K. Inaba and S. Suga, 
Phys. Rev. A {\bf 80}, 041602(R) (2009).

\bibitem{IPT}
A. Georges and W. Krauth, Phys. Rev. B {\bf 48}, 7167 (1993).

\bibitem{Capone}
M. Capone, C. Castellani, and M. Grilli, 
Phys. Rev. Lett. {\bf 88}, 126403 (2002).

\bibitem{Garg}
A. Garg, H. R. Krishnamurthy, and M. Randeria, 
Phys. Rev. B {\bf 72}, 024517 (2005).

\bibitem{Bauer}
J. Bauer, A. C. Hewson, and N. Dupuis, Phys. Rev. B {\bf 79}, 214518 (2009);
J. Bauer and A. C.  Hewson, Europhys. Lett. {\bf 85}, 27001 (2009).

\bibitem{Toschi}
A. Toschi, M. Capone, and C. Castellani,
Phys. Rev. B {\bf 72}, 235118 (2005);
A. Toschi, P. Barone, M. Capone, and C. Castellani,
New J. Phys. {\bf 7}, 7 (2005).

\bibitem{SFA}
M. Potthoff, Eur. Phys. J. B {\bf 36}, 335 (2003).

\bibitem{GeorgesZ}
A. Georges, G. Kotliar, and W. Krauth, Z. Phys. B {\bf 92}, 313 (1993).

\bibitem{Dao}
T.-L. Dao, M. Ferrero, A. Georges, M. Capone, and O. Parcollet,
Phys. Rev. Lett. {\bf 101}, 236405 (2008).

\bibitem{KogaQMC1}
A. Koga and P. Werner, J. Phys. Soc. Jpn. {\bf 79}, 064401 (2010);
J. Phys. Soc. Jpn. {\bf 79} 114401 (2010).

\bibitem{Hoshino}
S. Hoshino and Y. Kuramoto, arXiv:1309.5719.

\bibitem{KogaSF}
A. Koga and P. Werner, 
Phys. Rev. A {\bf 84}, 023638 (2011).


\bibitem{Haule}
K. Haule and G. Kotliar, Phys. Rev. B {\bf 76}, 104509 (2007).

\bibitem{Takemori}
N. Takemori and A. Koga, J. Phys. Soc. Jpn. {\bf 81}, 063002 (2012).

\bibitem{Bodensiek}
O. Bodensiek, R. Zitko, M. Vojta, M. Jarrell, and T. Pruschke, 
Phys. Rev. Lett. {\bf 110}, 146406 (2013).


\bibitem{Caffarel}
M. Caffarel and W. Krauth, Phys. Rev. Lett. {\bf 72}, 1545 (1994).  


\bibitem{Privitera}
A. Privitera, M. Capone, and C. Castellani, 
Phys. Rev. B {\bf 81}, 014523 (2010). 

\bibitem{NRG}
H. R. Krishna-murthy, J. W. Wilkins, and K. G. Wilson,
Phys. Rev. B {\bf 21}, 1003 (1980).

\bibitem{NRG_RMP} 
R. Bulla, T. Costi, and Th. Pruschke,
Rev. Mod. Phys. {\bf 80}, 395 (2008).

\bibitem{MEM1}
S. F. Gull, in {\it Maximum Entropy and Bayesian Methods in Science 
and Engineering}, ed. G. J. Erickson and C. R. Smith 
(Kluwer Academic, Dordrecht, 1988) p. 53;
J. Skilling(Kluwer Academic, Dordrecht, 1989) p. 45;
S. F. Gull, {\it ibid}. p. 53.

\bibitem{MEM2}
R. N. Silver, D. S. Sivia and J. E. Gubernatis, 
Phys. Rev. B {\bf 41}, 2380 (1990); 
J. E. Gubernatis, M. Jarrell, R. N. Silver and D. S. Sivia, 
Phys. Rev. B {\bf 44}, 6011 (1991).

\bibitem{MEM3}
W. F. Press, S. A. Teukolsky, W. T. Vetterling and B. R. Flannery, 
{\it Numerical Recipes} (Cambridge University Press, Cambridge, 1992) p. 809.


\bibitem{alps1.3}
A. F. Albuquerque, F. Alet, P. Corboz, P. Dayal, A. Feiguin, S.
Fuchs, L. Gamper, E. Gull, S. Gurtler, A. Honecker, R. Igarashi,
M. Korner, A. Kozhevnikov, A. Lauchli, S. R. Manmana, M.
Matsumoto, I. P. McCulloch, F. Michel, R. M. Noack, G.
Paw.owski, L. Pollet, T. Pruschke, U. Schollwock, S. Todo, S.
Trebst, M. Troyer, P. Werner, S. Wessel, 
J. Magn. Magn. Mater. {\bf 310}, 1187 (2007).

\end{thebibliography}
\end{document}